\documentclass[10pt]{article}
\usepackage{multicol}
\usepackage{graphicx}
\usepackage{amsmath}
\usepackage[a4paper]{geometry}

\begin{document}

\begin{center}
{\Large \bf Possible scenarios for single, double, or multiple
kinetic freeze-out\\ in high energy collisions}

\vskip1.0cm

Muhammad Waqas$^1$, Fu-Hu Liu$^{1,}${\footnote{E-mail:
fuhuliu@163.com; fuhuliu@sxu.edu.cn}}, Sakina Fakhraddin$^{2,3}$,
Magda A. Rahim$^{2,3}$

\vskip.25cm

{\small \it $^1$Institute of Theoretical Physics \& State Key
Laboratory of Quantum Optics and Quantum Optics Devices,\\ Shanxi
University, Taiyuan, Shanxi 030006, China

$^2$Physics Department, College of Science \& Arts in Riyadh
Al-Khabra,\\ Qassim University, Qassim, Kingdom of Saudi Arabia

$^3$Physics Department, Faculty of Science, Sana'a University,
P.O. Box 1247, Sana'a, Republic of Yemen}
\end{center}

\vskip1.0cm

{\bf Abstract:} Transverse momentum spectra of different types of
particles produced in mid-rapidity interval in central and
peripheral gold-gold (Au-Au) collisions, central and peripheral
deuteron-gold ($d$-Au) collisions, and inelastic (INEL) or
non-single-diffractive (NSD) proton-proton ($pp$) collisions at
the Relativistic Heavy Ion Collider (RHIC), as well as in central
and peripheral lead-lead (Pb-Pb) collisions, central and
peripheral proton-lead ($p$-Pb) collisions, and INEL or NSD $pp$
collisions at the Large Hadron Collider (LHC) are analyzed by the
blast-wave model with Boltzmann-Gibbs statistics. The model
results are largely consist with the experimental data in special
transverse momentum ranges measured by the PHENIX, STAR, ALICE,
and CMS Collaborations. It is showed that the kinetic freeze-out
temperature of emission source is dependent on particle mass,
which reveals the scenario for multiple kinetic freeze-out in
collisions at the RHIC and LHC. The scenario for single or double
kinetic freeze-out is not observed in this study.
\\



\vskip1.0cm
\begin{multicols}{2}

{\section{Introduction}}

Chemical freeze-out is an important intermediate stage in high
energy collisions. During this stage, the intra-nuclear collisions
among particles are inelastic, and ratios of different types of
particles remain invariant. Chemical freeze-out temperature
($T_{ch}$) is an important quantity which describes the excitation
degree of interacting system at the stage of chemical freeze-out.
It is important to know about the various $T_{ch}$ at the stage of
chemical freeze-out. Correspondingly, kinetic freeze-out is the
last stage, but not least one, in high energy collisions, in which
the intra-nuclear collisions among particles are elastic and the
transverse momentum distributions of different types of particles
are no longer changed. Kinetic freeze-out temperature ($T_0$ or
$T_{kin}$) is also an important quantity which describes the
excitation degree of the interacting system at the stage of
kinetic freeze-out. A natural question is that how many different
$T_0$ are there at the stage of kinetic freeze-out.

Generally, the kinetic freeze-out happens later than or
simultaneously with the chemical freeze-out. This renders that
$T_0$ is smaller than or equal to $T_{ch}$. According to the
thermal and statistical model [1--4], with the increase of the
collision energy, the single $T_{ch}$ in central nucleus-nucleus
collisions increases from a few GeV to above 10 GeV, and then
saturates in an energy range more than dozens of GeV. The maximum
$T_{ch}$ at the top Relativistic Heavy Ion Collider (RHIC) and the
Large Hadron Collider (LHC) energies is about 160 MeV, though
there is a slightly increase from the top RHIC energy to that of
LHC. Meanwhile, $T_{ch}$ in central nucleus-nucleus collisions is
slightly larger than that in peripheral nucleus-nucleus
collisions. The properties of $T_{ch}$ is well acknowledged in the
community.

However, the situation of $T_0$ is more complex. Firstly, with
increasing the collision energy, although $T_0$ in central
collisions increases initially over the energy range from a few
GeV to above 10 GeV, the subsequent tendency can be saturated,
increscent, or decrescent. It is interesting to know the correct
subsequent tendency. Secondly, $T_0$ in central collisions can be
larger than, equal to, or smaller than that in peripheral
collisions. It is important to figure out which $T_0$ is larger.
Thirdly, $T_0$ can possibly give single, double, or multiple
values for the emissions of different types of particles in the
given collisions. Which freeze-out scenario is correct also
interests us.

It is crucial but also challenging to solve all the above three
issues. In particular, the first issue needs to study the
excitation function of $T_0$, which needs the collection and
analysis of many experimental data. We shall do it in other work.
The second issue is studied in our very recent work [5] which
shows a slightly larger $T_0$ in central collisions if a non-zero
transverse flow velocity ($\beta_T$) is used for peripheral
collisions. Considering the low multiplicity of the final state
particles, peripheral collisions at the RHIC are similar to
inelastic (INEL) or non-single-diffractive (NSD) proton-proton
($pp$) collisions at the LHC. Given the low frequency of the
cascade collisions in both participant nucleons, peripheral
collisions are also similar to INEL or NSD $pp$ collisions. In
fact, $pp$ collisions also show collective expansion [6] and then
reveal non-zero $\beta_T$. For the third issue, we have to extract
$T_0$ from the transverse momentum ($p_T$) spectra of different
types of particles. As an accompanying result, $\beta_T$ can be
obtained, which also shows complex situation as $T_0$.

Several methods can be used to calculate $T_0$ and $\beta_T$. In
the present work, we shall use the blast-wave model with
Boltzmann-Gibbs statistics [7--9] to extract $T_0$ and $\beta_T$
from $p_T$ spectra of different types of particles produced in
mid-rapidity interval in central and peripheral gold-gold (Au-Au)
collisions, central and peripheral deuteron-gold ($d$-Au)
collisions, and INEL or NSD $pp$ collisions at the RHIC, as well
as in central and peripheral lead-lead (Pb-Pb) collisions, central
and peripheral proton-lead ($p$-Pb) collisions, and INEL or NSD
$pp$ collisions at the LHC. The contribution of soft excitation
process is included, while the contribution of hard scattering
process is not excluded, if available in low $p_T$ region. The
model results are compared with the experimental data measured by
the PHENIX [10--14], STAR [15--18], ALICE [19--27], and CMS [28]
Collaborations.

The remainder of this paper is structured as follows. The method
and formalism are shortly described in Section 2. Results and
discussions are given in Section 3. In Section 4, we summarize our
main observations and conclusions.
\\

{\section{The method and formalism}}

A few methods can be used to extract $T_0$ and $\beta_T$,
including but not limited to, i) the blast-wave model with
Boltzmann-Gibbs statistics [7--9], ii) the blast-wave model with
Tsallis statistics [29], iii) an alternative method using the
Boltzmann distribution [8, 30--36], iv) the alternative method
using the Tsallis distribution [37, 38]. It should be noted that,
in the alternative method, $T_0$ is the intercept in the linear
relation $T-m_0$, where $T$ is the effective temperature which
includes the contributions of thermal motion and flow effect, and
$m_0$ is the rest mass; and $\beta_T$ is the slope in the linear
relation $\langle p_T \rangle -\overline{m}$, where $\langle p_T
\rangle$ is the mean transverse momentum and $\overline{m}$ is the
mean moving mass (i.e. the mean energy).

Our very recent work [5] confirms that the above methods are
harmonious. Among these methods, the first one is the most direct
and has fewer parameters, though it is revised in different ways
and applied to other quantities [39--43]. In the present work, we
have used the first method i.e. the blast-wave model with
Boltzmann-Gibbs statistics, to extract $T_0$ and $\beta_T$. Other
methods will not be used due to their coherence [5]. As an
application of the blast-wave model with Boltzmann-Gibbs
statistics, we shall only give a short representation of its
formalism in the original form. The discussions on its various
revisions and applications to derive other quantities are beyond
the focus of the present work. We will not discuss them further.

According to refs. [7--9], the blast-wave model with
Boltzmann-Gibbs statistics gives the following $p_T$ distribution
\begin{align}
f_1(p_T)=&\frac{1}{N}\frac{dN}{dp_T} =C p_T m_T \int_0^R rdr \nonumber\\
& \times I_0 \bigg[\frac{p_T \sinh(\rho)}{T_0} \bigg] K_1
\bigg[\frac{m_T \cosh(\rho)}{T_0} \bigg],
\end{align}
where $N$ is the number of particles, $C$ is the normalized
constant, $m_T=\sqrt{p_T^2+m_0^2}$ is the transverse mass, $I_0$
and $K_1$ are the modified Bessel functions of the first and
second kinds respectively, $\rho= \tanh^{-1} [\beta(r)]$ is the
boost angle, $\beta(r)= \beta_S(r/R)^{n_0}$ is a self-similar flow
profile, $\beta_S$ is the flow velocity on the surface, $r/R$ is
the relative radial position in the thermal source, and $n_0$ is a
free parameter [7].

One has the relation, $\beta_T=(2/R^2)\int_0^R r\beta(r)dr =
2\beta_S/(n_0+2)$, between $\beta_T$ and $\beta(r)$. In the
present work, we use $n_0=2$ from ref. [7], which results in
$\beta_T=0.5\beta_S$. Because the maximum value of $\beta_S$ is
$1c$, the maximum value of $\beta_T$ is $0.5c$. In other work [29]
which concerns the blast-wave model with Tsallis statistics,
$n_0=1$, which results in $\beta_T=(2/3)\beta_S$. Thus, the
maximum $\beta_T$ is $(2/3)c$. Another work [19] uses $n_0$ to be
a non-integer from that less than 1 to above 2, which corresponds
to the centrality from center to periphery. This can lead to a
large variation in $\beta_T$. As a not too sensitive quantity, the
selection of $n_0$ is flexible. Although different $n_0$ can be
used to fit $p_T$ spectrum, it impacts generally $T_0$. Meanwhile,
$\beta_T$ also impacts $T_0$.

Generally, there are mainly two processes, the soft excitation
process and the hard scattering process, in the contributions for
$p_T$ spectrum. The soft excitation process contributes $p_T$
spectrum in a narrow range. The hard excitation process
contributes $p_T$ spectrum in a wide range. The blast-wave model
with Boltzmann-Gibbs statistics and other methods mentioned above
describe only the contribution of soft excitation process. For the
contribution of hard scattering process, we can use an inverse
power-law [44--46], i.e. the Hagedorn function [47, 48]
\begin{align}
f_2(p_T)=\frac{1}{N}\frac{dN}{dp_T}= Ap_T \bigg( 1+\frac{p_T}{
p_0} \bigg)^{-n},
\end{align}
where $p_0$ and $n$ are free parameters and $A$ is the
normalization constant related to the free parameters. The
Hagedorn function has three revisions [49--55] which will not be
discussed here because they are out of the scope of this work. To
describe a wide $p_T$ spectrum, we can use a superposition of the
two contributions.

We have two methods to superpose the two functions, $f_1(p_T)$ and
$f_2(p_T)$. Considering the continuities of the two functions from
0 to each maximum, we have
\begin{align}
f_0(p_T)=\frac{1}{N}\frac{dN}{dp_T}=kf_1(p_T)+(1-k)f_2(p_T),
\end{align}
where $k$ denotes the contribution fraction of the first
component, $f_1(p_T)$, the soft excitation process. According to
Hagedorn's model [47], we can also use the usual step function
\begin{align}
f_0(p_T)&=\frac{1}{N}\frac{dN}{dp_T} \nonumber\\ &=A_1
\theta(p_1-p_T) f_1(p_T) + A_2 \theta(p_T-p_1)f_2(p_T),
\end{align}
to superpose the two functions. Here $A_1$ and $A_2$ are constants
which result in the two functions to be equal at $p_T=p_1$.

The first superposition (Eq. (3)) has been used in our recent work
[5]. In a low $p_T$ region, there is an entanglement between
$f_1(p_T)$ and $f_2(p_T)$ in the first superposition, though the
contribution of $f_1(p_T)$ dominates in most cases. We will use
the second superposition (Eq. (4)) in the present work. In the
second superposition, there is no entanglement between $f_1(p_T)$
and $f_2(p_T)$ in different $p_T$ regions. In particular,
$f_1(p_T)$ contributes only in a low $p_T$ region, and $f_2(p_T)$
contributes only in a high $p_T$ region. A narrow $p_T$ range such
as 0--5 GeV/$c$ is enough to extract $T_0$ and $\beta_T$ by the
second superposition. In the case of encountering a wide $p_T$
spectrum, we can cut only the range of 0--5 GeV/$c$ for analysis
in the second superposition. In fact, even a range of 0--$2\sim3$
GeV/$c$ is already wide enough in some cases. In the calculation,
only $f_1(p_T)$ is used in the low $p_T$ region which is a special
region as wide as possible, and $f_2(p_T)$ which can be used in
the high $p_T$ region is not used, as it is being beyond the focus
of the present work.

Both the superpositions in Eqs. (3) and (4) are normalized to 1
because they are the probability density functions. We need a
normalization constant, $N_0$, in the fitted process if the
experimental data are not normalized to 1. In many cases, the
experimental data presented in literature are not in the form of
probability density function. Thus, we need an appropriate
transformation for the superpositions so that we can compare them
with the experimental data. For example, for the main three forms
of experimental $p_T$ spectra, $(1/2\pi p_T)d^2N/dp_Tdy$,
$d^2N/dp_Tdy$, and $dN/dp_T$, one can use $(1/2\pi
p_T)N_0f_0(p_T)/dy$, $N_0f_0(p_T)/dy$, and $N_0f_0(p_T)$ to fit
them, respectively. The purpose of the present work is to extract
$T_0$ and $\beta_T$. The meaning and value of $N_0$ are less
important, though $N_0$ can be obtained at the volley. Instead,
the ratio of $N_0$ for different types of particles in the same
form of spectra measured in the same experimental condition have
important meanings.
\\

{\section{Results and discussion}}

Figure 1 presents the transverse momentum spectra, $(1/2\pi
p_T)d^2N/dp_Tdy$, of different types of particles produced in
mid-rapidity interval in Au-Au collisions at center-of-mass energy
$\sqrt{s_{NN}}=200$ GeV at the RHIC. The symbols represent the
experimental data measured by the PHENIX [10] and STAR [15, 16]
Collaborations. The curves are our fitted results by using the
blast-wave model with Boltzmann-Gibbs statistics, Eq. (1). In the
bottom part of each panel, the results of data/fit are presented
to monitor the difference between fit and data. The left and right
panels are the results corresponding to central (0--5\%
centrality) and peripheral (60--92\% or 60--80\% centrality)
collisions respectively. The upper, middle, and lower panels are
the results corresponding to positive pions ($\pi^+$), positive
kaons ($K^+$), and protons ($p$); negative pions ($\pi^-$),
negative kaons ($K^-$), and anti-protons ($\bar p$); as well as
neutral $\phi$ mesons and negative $\it \Xi^-$ baryons,
respectively. In each fitting, the method of least squares is
used, in which the minimum $\chi^2$ is used to determine various
parameters, where $\chi^2=\sum_i[(f_i|_{\rm exp}-f_i|_{\rm
th})^2/\sigma_i^2]$, $f_i|_{\rm exp}$ denotes the experimental
probability, $f_i|_{\rm th}$ denotes the theoretical probability,
and $\sigma_i$ denotes the experimental error, corresponding to
the $i$th data point. The values of free parameters ($T_0$ and
$\beta_T$), normalization constant ($N_0$), $\chi^2$, and degrees
of freedom (dof) are listed in Table 1. In most cases, the
confidence levels of fittings are 95\%. In a few cases, these
values are 90\%. To avoid trivialness, the concrete confidence
levels are not listed in the table one by one. One can see that
the model results describe approximately the experimental data in
special transverse momentum ranges in large collision system
measured at the RHIC by the PHENIX and STAR Collaborations. The
special transverse momentum ranges for some particles in
peripheral Au-Au collisions are only in 0--$2\sim3$ GeV/$c$. In
other cases, the particles are expected to have wider special
transverse momentum ranges.

\begin{figure*}[htbp]
\vskip-1cm
\begin{center}
\includegraphics[width=16.cm]{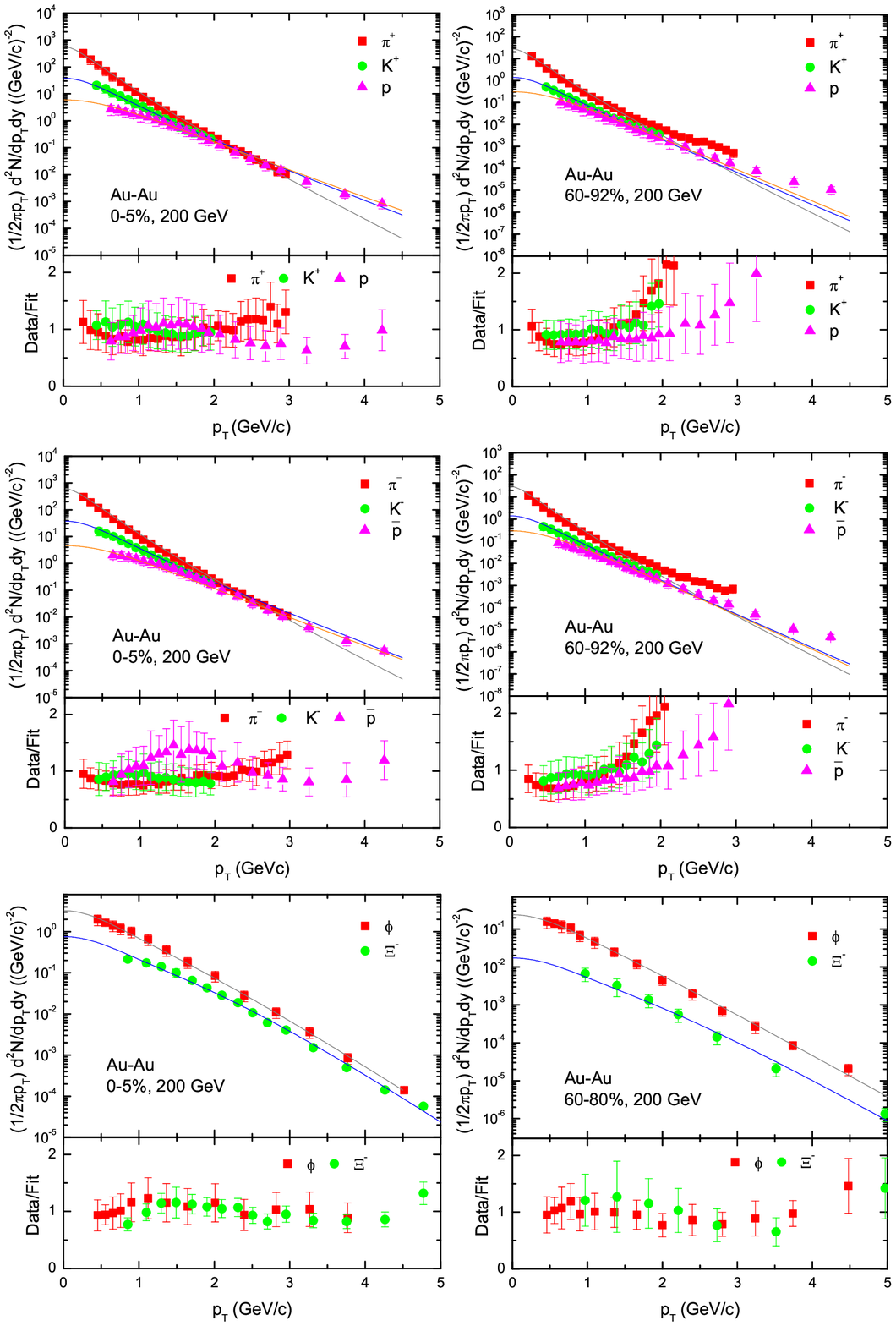}
\end{center}
{\small Fig. 1. Transverse momentum spectra of different types of
particles produced in Au-Au collisions at $\sqrt{s_{NN}}=200$ GeV
at the RHIC. The symbols represent the experimental data measured
by the PHENIX [10] and STAR [15, 16] Collaborations and the curves
are our fitted results by using the blast-wave model with
Boltzmann-Gibbs statistics, Eq. (1). Following each panel, the
results of data/fit are presented. Left: for central collisions
(0--5\% centrality); Right: for peripheral collisions (60--92\% or
60--80\% centrality). Upper: for $\pi^+$, $K^+$, and $p$ with
$|\eta|<0.35$ [10]; Middle: for $\pi^-$, $K^-$, and $\bar p$ with
$|\eta|<0.35$ [10]; Lower: for $\phi$ with $|y|<0.5$ [15] and $\it
\Xi^-$ with $|y|<0.75$ [16].}
\end{figure*}

\begin{table*}[!htb]
{\small Table 1. Values of $T_0$, $\beta_T$, $N_0$, $\chi^2$, and
dof corresponding to the curves in Figs. 1--6.} \vspace{-.30cm}
{\scriptsize
\begin{center}
\begin{tabular}{ccccccc}\\ \hline\hline
Figure & Centrality & Particle & $T_0$ (GeV) & $\beta_T$ ($c$) &
 $N_0$ & $\chi^2$/dof \\ \hline
Fig. 1  & 0--5\%   & $\pi^+$  & $0.152\pm0.001$ & $0.344\pm0.001$ & $47.60\pm 2.40$ & 13/25\\
Au-Au   &          & $K^+$    & $0.120\pm0.002$ & $0.440\pm0.003$ & $7.70\pm0.30$   &  6/13\\
200 GeV &          & $p$      & $0.220\pm0.002$ & $0.321\pm0.002$ & $2.66\pm0.10$   & 10/19\\
        &          & $\pi^-$  & $0.140\pm0.001$ & $0.368\pm0.001$ & $ 49.50\pm2.00$ & 30/25\\
        &          & $K^-$    & $0.120\pm0.002$ & $0.335\pm0.002$ & $ 7.70\pm0.25$  & 10/13\\
        &          & $\bar p$ & $0.210\pm0.001$ & $0.333\pm0.002$ & $1.97\pm0.10$   & 17/19\\
        &          & $\phi$   & $0.125\pm0.002$ & $0.400\pm0.002$ & $1.20\pm0.01$   &  2/11\\
        &          & $\Xi^-$  & $0.120\pm0.001$ & $0.395\pm0.001$ & $0.56\pm0.03$   & 15/12\\
\cline{2-7}
        & 60--92\% & $\pi^+$  & $0.115\pm0.003$ & $0.347\pm0.003$ & $1.70\pm0.05$   & 74/25\\
        &          & $K^+$    & $0.090\pm0.002$ & $0.400\pm0.003$ & $0.20\pm0.01$   & 11/13\\
        &          & $p$      & $0.130\pm0.002$ & $0.338\pm0.003$ & $0.090\pm0.003$ & 14/19\\
        &          & $\pi^-$  & $0.115\pm0.002$ & $0.372\pm0.003$ & $1.80\pm0.08$   & 46/25\\
        &          & $K^-$    & $0.090\pm0.003$ & $0.407\pm0.004$ & $0.20\pm0.01$   &  4/13\\
        &          & $\bar p$ & $0.138\pm0.003$ & $0.308\pm0.004$ & $0.090\pm0.004$ & 31/19\\
\cline{2-7}
        & 60--80\% & $\phi$   & $0.150\pm0.001$ & $0.380\pm0.001$ & $0.095\pm0.002$ &  4/11\\
        &          & $\Xi^-$  & $0.135\pm0.002$ & $0.390\pm0.002$ & $0.014\pm0.002$ &  4/4\\
\hline
Fig. 2  & 0--20\%  & $\pi^+$  & $0.120\pm0.001$ & $0.439\pm0.001$ & $1.00\pm0.10$     & 21/21\\
$d$-Au  &          & $K^+$    & $0.280\pm0.002$ & $0.200\pm0.002$ & $0.11\pm0.01$     & 12/18\\
200 GeV &          & $p$      & $0.202\pm0.002$ & $0.340\pm0.002$ & $0.060\pm0.005$   &  6/21\\
        &          & $\pi^-$  & $0.120\pm0.001$ & $0.442\pm0.001$ & $0.87\pm0.07$     &  9/21\\
        &          & $K^-$    & $0.270\pm0.002$ & $0.204\pm0.005$ & $0.11\pm0.01$     &  2/18\\
        &          &$\bar p$  & $0.208\pm0.002$ & $0.330\pm0.002$ & $0.042\pm0.002$   & 22/21\\
        &          & $\eta$   & $0.125\pm0.001$ & $0.436\pm0.001$ & $ 0.065\pm0.004$  &  9/3\\
        &          & $\phi$   & $0.110\pm0.002$ & $0.420\pm0.002$ & $ 0.015\pm0.001$  &  7/5\\
\cline{2-7}
        & 60--88\% & $\pi^+$  & $0.125\pm0.002$ & $0.388\pm0.002$ & $0.30\pm0.03$     &  5/21\\
        &          & $K^+$    & $0.265\pm0.002$ & $0.140\pm0.003$ & $0.030\pm0.002$   & 25/18\\
        &          & $p$      & $0.177\pm0.001$ & $0.290\pm0.002$ & $0.015\pm0.001$   & 25/21\\
        &          & $\pi^-$  & $0.117\pm0.002$ & $0.400\pm0.002$ & $0.30\pm0.01$     & 21/21\\
        &          & $K^-$    & $0.251\pm0.001$ & $0.155\pm0.003$ & $0.030\pm0.001$   & 10/18\\
        &          & $\bar p$ & $0.199\pm0.002$ & $0.285\pm0.003$ & $0.010\pm0.001$   & 11/21\\
        &          & $\eta$   & $0.110\pm0.003$ & $0.445\pm0.002$ & $0.018\pm0.002$   &  1/3\\
        &          & $\phi$   & $0.155\pm0.004$ & $0.375\pm0.003$ & $0.0027\pm0.0004$ &  6/5\\
\hline
Fig. 3  & $-$      & $\pi^+$  & $0.114\pm0.002$ & $0.402\pm0.002$ & $3.78\pm0.14$      & 71/24\\
$pp$    &          & $K^+$    & $0.200\pm0.002$ & $0.200\pm0.003$ & $0.43\pm0.02$      & 24/13\\
200 GeV &          & $p$      & $0.145\pm0.002$ & $0.326\pm0.002$ & $0.17\pm0.01$      & 120/24\\
        &          & $\pi^-$  & $0.125\pm0.002$ & $0.383\pm0.002$ & $3.96\pm0.49$      & 54/24\\
        &          & $K^-$    & $0.197\pm0.003$ & $0.190\pm0.003$ & $0.44\pm0.04$      & 14/13\\
        &          & $\bar p$ & $0.144\pm0.001$ & $0.322\pm0.001$ & $0.13\pm0.01$      & 232/30\\
        &          & $\phi$   & $0.112\pm0.002$ & $0.380\pm0.002$ & $0.0031\pm0.0003$  & 31/10\\
        &          & $\Xi^-$  & $0.180\pm0.002$ & $0.298\pm0.005$ & $0.00042\pm0.00005$& 12/8\\
\hline
Fig. 4   & 0--5\%  & $\pi^++\pi^-$ & $0.135\pm0.003$ & $0.432\pm0.003$ & $251.00\pm20.00$ & 5/38\\
Pb-Pb    &         & $K^++K^-$     & $0.289\pm0.003$ & $0.264\pm0.002$ & $33.64\pm1.00$   & 4/33\\
2.76 TeV &         & $p+\bar p$    & $0.443\pm0.002$ & $0.098\pm0.003$ & $10.50\pm0.50$   & 6/34\\
\cline{2-7}
         & 0--10\% & $\phi$        & $0.180\pm0.002$ & $0.370\pm0.002$ & $24.39\pm2.00$   & 1/5\\
         &         & $\it \Xi^-$   & $0.320\pm0.002$ & $0.305\pm0.002$ & $3.80\pm0.03$    & 252/19\\
\cline{2-7}
         & 70--80\%& $\pi^++\pi^-$ & $0.133\pm0.002$ & $0.420\pm0.002$ & $5.50\pm0.30$    & 10/38\\
         &         & $K^++K^-$     & $0.210\pm0.002$ & $0.347\pm0.002$ & $0.77\pm0.03$    & 2/33\\
         &         & $p+\bar p$    & $0.222\pm0.001$ & $0.355\pm0.001$ & $0.27\pm0.01$    & 16/34\\
\cline{2-7}
         & 60--80\%& $\phi$        & $0.199\pm0.007$ & $0.405\pm0.007$ & $0.37\pm0.02$    & 5/5\\
         &         & $\it \Xi^-$   & $0.183\pm0.009$ & $0.382\pm0.002$ & $0.13\pm0.02$    & 13/17\\
\hline
Fig. 5   & 0--5\%   & $\pi^++\pi^-$             & $0.119\pm0.001$ & $0.469\pm0.001$ & $7.23\pm0.04$     & 45/38\\
$p$-Pb   &          & $K^++K^-$                 & $0.293\pm0.003$ & $0.313\pm0.004$ & $0.98\pm0.04$     & 13/28\\
5.02 TeV &          & $p+\bar p$                & $0.265\pm0.002$ & $0.393\pm0.002$ & $0.34\pm0.01$     & 12/36\\
         &          & $\phi$                    & $0.208\pm0.002$ & $0.435\pm0.001$ & $0.17\pm0.01$     & 16/11\\
         &          & $\it(\Xi^-+\bar \Xi^+)$/2 & $0.300\pm0.007$ & $0.390\pm0.005$ & $0.0099\pm0.0005$ & 16/12\\
\cline{2-7}
         & 60--80\% & $\pi^++\pi^-$             & $0.120\pm0.003$ & $0.453\pm0.003$ & $1.45\pm0.10$     & 69/36\\
         &          & $K^++K^-$                 & $0.232\pm0.002$ & $0.329\pm0.004$ & $0.19\pm0.01$     & 23/28\\
         &          & $p+\bar p$                & $0.215\pm0.002$ & $0.365\pm0.002$ & $0.10\pm0.01$     & 107/36\\
         &          & $\phi$                    & $0.170\pm0.003$ & $0.440\pm0.002$ & $0.033\pm0.001$   & 15/11\\
         &          & $\it(\Xi^-+\bar \Xi^+)$/2 & $0.230\pm0.001$ & $0.399\pm0.002$ & $0.0017\pm0.0001$ & 4/12\\
\hline
Fig. 6 & $-$ & $\pi^++\pi^-$ & $0.130\pm0.002$ & $0.430\pm0.002$ & $4.35\pm0.10$   & 206/35\\
$pp$   &     & $K^++K^-$     & $0.120\pm0.003$ & $0.458\pm0.002$ & $0.58\pm0.02$   & 374/43\\
7 TeV  &     & $p+\bar p$    & $0.204\pm0.003$ & $0.350\pm0.004$ & $0.25\pm0.01$   & 678/39\\
       &     & $\phi$        & $0.130\pm0.002$ & $0.430\pm0.001$ & $0.031\pm0.003$ & 108/22\\
       &     & $\it\Xi^-$    & $0.280\pm0.004$ & $0.320\pm0.005$ & $0.085\pm0.003$ & 138/18\\
\hline
\end{tabular}%
\end{center}}
\end{table*}

Figure 2 is the same as Fig. 1, but it shows the transverse
momentum spectra of different types of particles produced in
mid-rapidity interval in $d$-Au collisions at $\sqrt{s_{NN}}=200$
GeV. The symbols represent the experimental data measured by the
PHENIX Collaboration [11--13], where the spectra of neutral $\eta$
mesons and $\phi$ are cut at 5 GeV/$c$ due to a wide range being
unnecessary for the extractions of $T_0$ and $\beta_T$. The left
and right panels are the results corresponding to central (0--20\%
centrality) and peripheral (60--88\% centrality) collisions
respectively. The upper, middle, and lower panels are the results
corresponding to $\pi^+$, $K^+$, and $p$; $\pi^-$, $K^-$, and
$\bar p$; as well as $\eta$ and $\phi$, respectively. Figure 3 is
also the same as Fig. 1, but it shows the transverse momentum
spectra with different expression in mid-rapidity interval in $pp$
collisions at center-of-mass energy $\sqrt{s}=200$ GeV, where $E$,
$\sigma$, and $N_{\rm ev}$ on the vertical axis denote the energy,
cross section, and event number, respectively. The symbols
represent the experimental data measured by the PHENIX [14] and
STAR [17, 18] Collaborations. The left-upper, right-upper, and
lower panels are the results corresponding to $\pi^+$, $K^+$, and
$p$ in INEL events; $\pi^-$, $K^-$, and $\bar p$ in INEL events;
as well as $\phi$ and $\it \Xi^-$ in NSD events, respectively. One
can see that the model results describe approximately the
experimental data in special transverse momentum ranges in small
collision system measured at the RHIC by the PHENIX and STAR
Collaborations. The special transverse momentum ranges for some
particles in peripheral $d$-Au collisions are only in 0--$2\sim3$
GeV/$c$ or a little more. In other cases, the particles are
expected to have wider special transverse momentum ranges.

\begin{figure*}[htbp]
\vskip-1cm
\begin{center}
\includegraphics[width=16.cm]{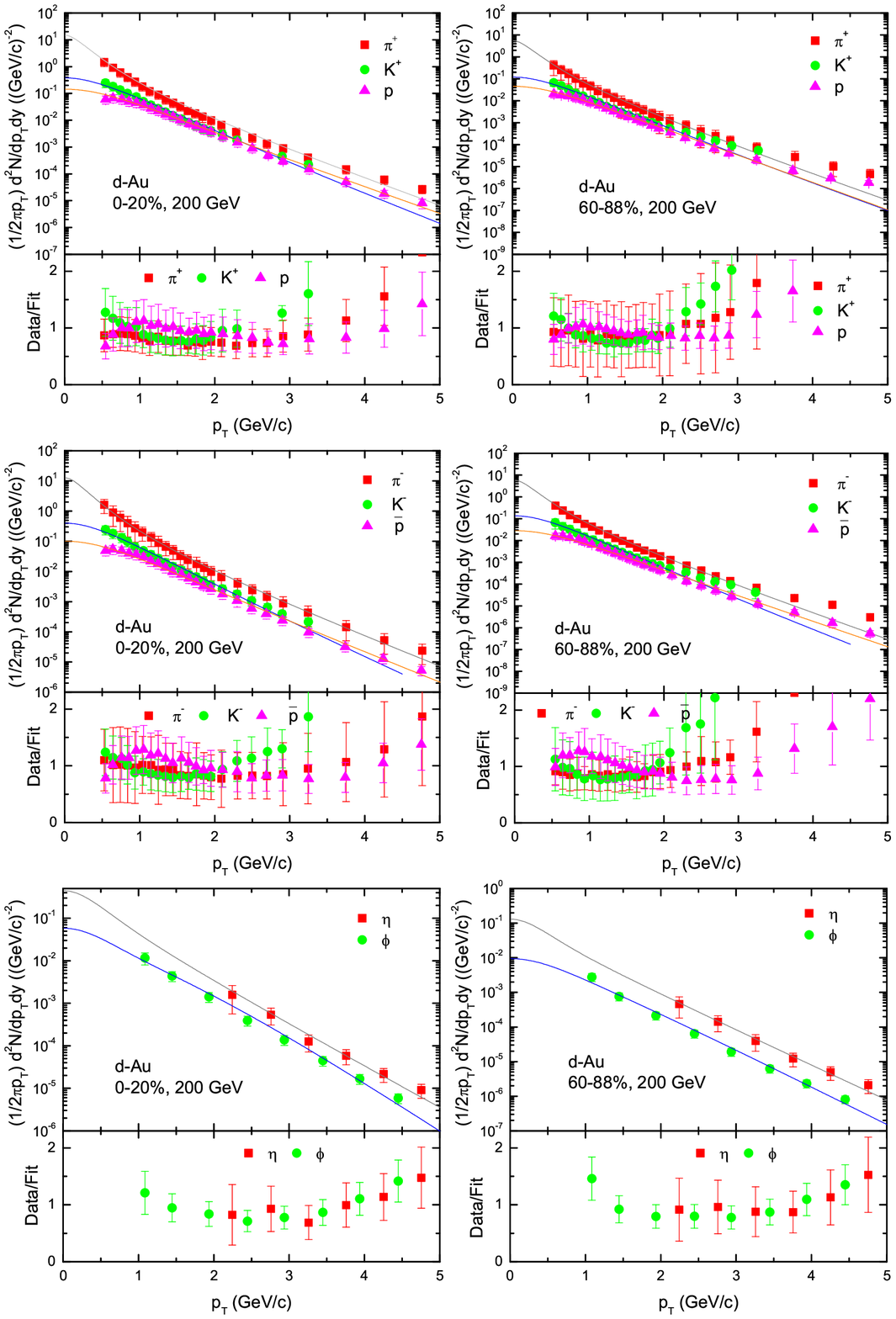}
\end{center}
Fig. 2. Same as Fig. 1, but showing the spectra in $d$-Au
collisions at $\sqrt{s_{NN}}=200$ GeV. The symbols represent the
experimental data measured by the PHENIX Collaboration [11--13],
where the spectra of $\eta$ and $\phi$ are cut at 5 GeV/$c$ due to
a wide range being unnecessary for the extractions of $T_0$ and
$\beta_T$. Left: for central collisions (0--20\% centrality);
Right: for peripheral collisions (60--88\% centrality). Upper: for
$\pi^+$, $K^+$, and $p$ with $|\eta|<0.35$ [11]; Middle: for
$\pi^-$, $K^-$, and $\bar p$ with $|\eta|<0.35$ [11]; Lower: for
$\eta$ [12] and $\phi$ [13] with $|\eta|<0.35$.
\end{figure*}

\begin{figure*}[!htb]
\begin{center}
\includegraphics[width=16.cm]{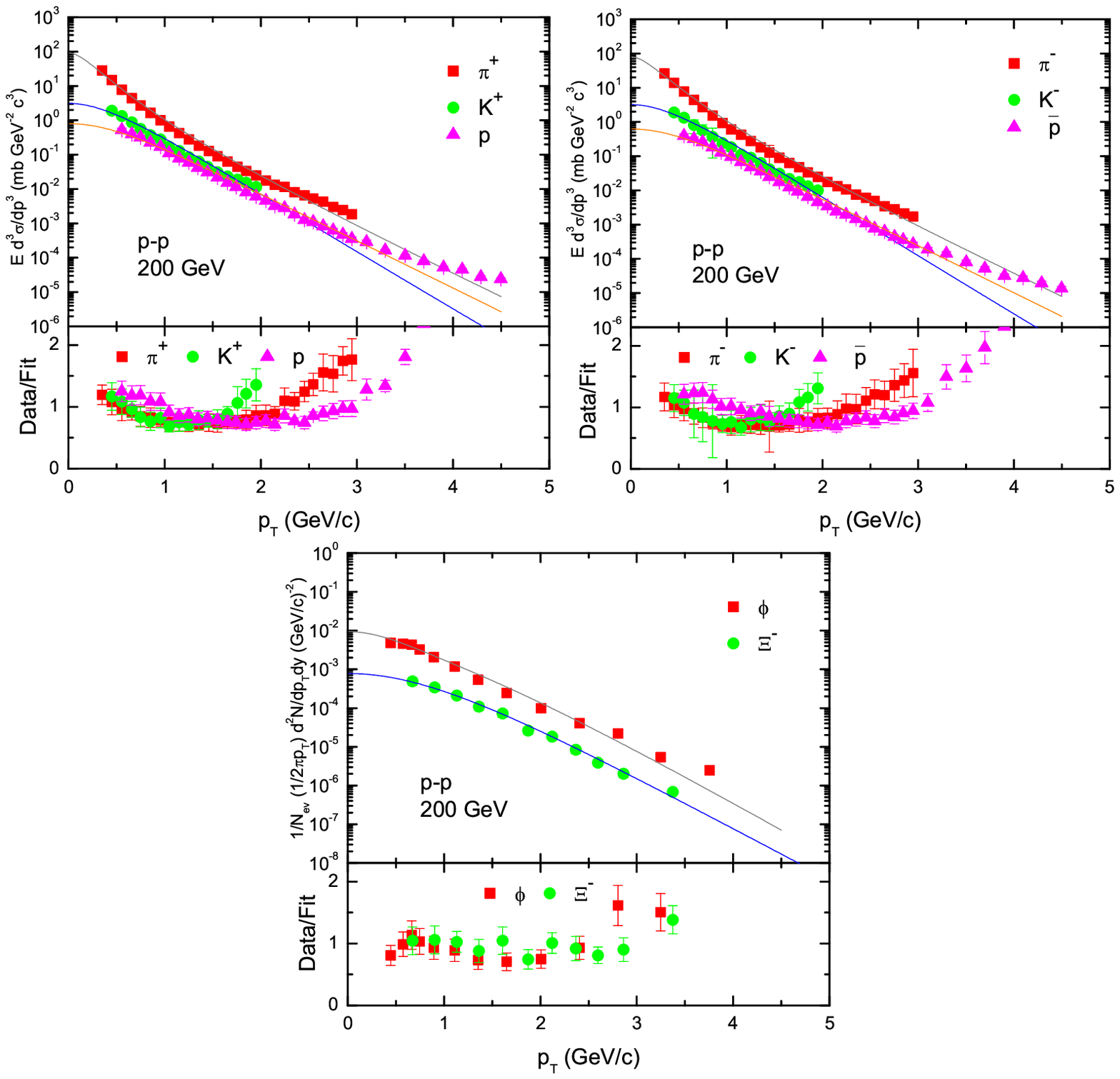}
\end{center}
Fig. 3. Same as Fig. 1, but showing the spectra with different
expression in $pp$ collisions at $\sqrt{s}=200$ GeV, where $E$,
$\sigma$, and $N_{\rm ev}$ denote the energy, cross section, and
event number, respectively, and $N_{\rm ev}$ is usually omitted.
The symbols represent the experimental data measured by the PHENIX
[14] and STAR [17, 18] Collaborations in INEL and NSD events
respectively. Upper: for $\pi^+$, $K^+$, and $p$ (left), as well
as $\pi^-$, $K^-$, and $\bar p$ (right) with $|\eta|<0.35$ in INEL
events [14]; Lower: for $\phi$ [17] and $\it \Xi^-$ [18] with
$|y|<0.5$ in NSD events.
\end{figure*}

The transverse momentum spectra of different types of particles
produced in mid-rapidity interval in Pb-Pb collisions at
$\sqrt{s_{NN}}=2.76$ TeV at the LHC are displayed in Fig. 4, where
different expressions of $\phi$ and $\it \Xi^-$ spectra are used,
and different amounts marked in the panel are used to scale the
spectra of $\pi^++\pi^-$ and $K^++K^-$. The symbols represent the
experimental data measured by the ALICE Collaboration [19--22],
where the spectrum of $\it \Xi^-$ is cut at 5 GeV/$c$. The curves
are the results fitted by us using the blast-wave model with
Boltzmann-Gibbs statistics, Eq. (1). Following each panel, the
results of data/fit are presented. The left and right panels are
the results corresponding to central (0--5\% or 0--10\%
centrality) and peripheral (70--80\% or 60--80\% centrality)
collisions respectively. The upper and lower panels are the
results corresponding to $\pi^++\pi^-$, $K^++K^-$, and $p+\bar p$;
as well as $\phi$ and $\it \Xi^-$, respectively. The values of
$T_0$, $\beta_T$, $N_0$, $\chi^2$, and dof are listed in Table 1.
One can see that the model results describe approximately the
experimental data in special transverse momentum ranges in large
collision system measured at the LHC by the ALICE Collaboration.
The special transverse momentum ranges in Pb-Pb collisions are not
observed in the available data ranges.

\begin{figure*}[!htb]
\begin{center}
\includegraphics[width=16.cm]{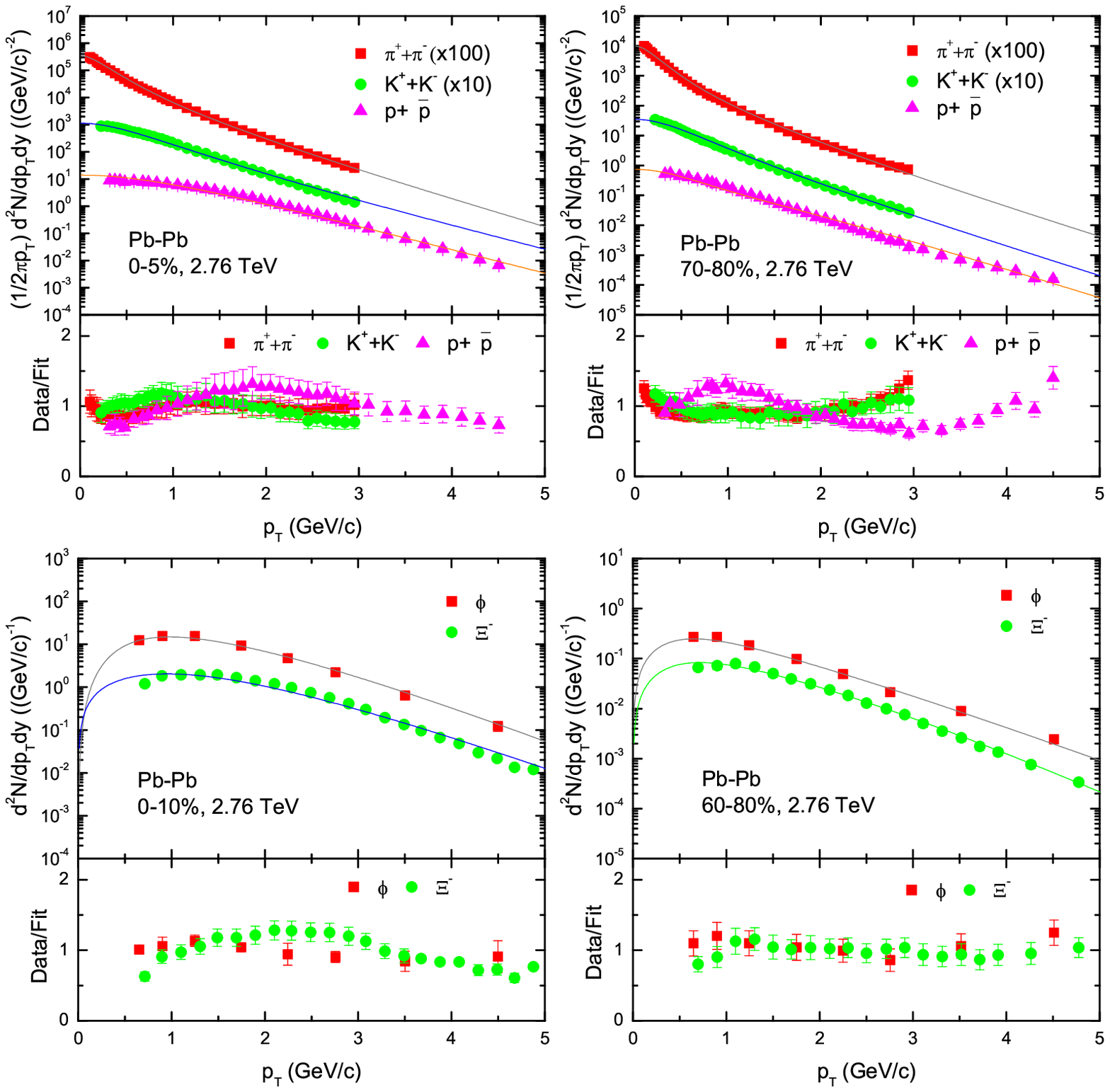}
\end{center}
Fig. 4. Transverse momentum spectra of different types of
particles produced in Pb-Pb collisions at $\sqrt{s_{NN}}=2.76$ TeV
at the LHC, where different expression of $\phi$ and $\Xi^-$
spectra is used. The symbols represent the experimental data
measured by the ALICE Collaboration [19--22], where the spectrum
of $\it \Xi^-$ is cut at 5 GeV/$c$. The curves are the results
fitted by us using the blast-wave model with Boltzmann-Gibbs
statistics, Eq. (1). Left: for central collisions (0--5\% or
0--10\% centrality); Right: for peripheral collisions (70--80\% or
60--80\% centrality). Upper: for $\pi^++\pi^-$, $K^++K^-$, and
$p+\bar p$ with $|y|<0.5$, where different amounts marked in the
panel are used to scale the spectra [19, 20]; Lower: for $\phi$
with $|y|<0.5$ [21] and $\it \Xi^-$ with $|y|<0.5$ at $p_T>1.8$
GeV/$c$ and with $|y|<0.3$ at $p_T<1.8$ GeV/$c$ [22].
\end{figure*}

Figure 5 is the same as Fig. 4, but it shows the transverse
momentum spectra of different types of particles produced in
mid-rapidity interval in $p$-Pb collisions at $\sqrt{s_{NN}}=5.02$
TeV. The symbols represent the experimental data measured by the
ALICE Collaboration [23--25], where the spectra of $\phi$ and
$\it(\Xi^-+\bar\Xi^+)/$2 are cut at 5 GeV/$c$, where $\it
\bar\Xi^+$ is anti-particle of positive $\it \Xi^+$ baryon. The
left and right panels are the results corresponding to central
(0--5\% centrality) and peripheral (60--80\% centrality)
collisions respectively. The upper and lower panels are the
results corresponding to $\pi^++\pi^-$, $K^++K^-$, and $p+\bar p$;
as well as $\phi$ and $\it(\Xi^-+\bar\Xi^+)/$2, respectively.
Figure 6 is the same as Fig. 4, too, but it showing the transverse
momentum spectra with different expression in mid-rapidity
interval in $pp$ collisions at $\sqrt{s}=7$ TeV, where $N_{\rm
INEL}$ and $N_{\rm NSD}$ on the vertical axis denote the numbers
of INEL and NSD events, respectively. The symbols represent the
experimental data measured by the ALICE [26, 27] and CMS [28]
Collaborations, where the spectra of $K^++K^-$, $p+\bar p$,
$\phi$, and $\it \Xi^-$ are cut at 5 GeV/$c$. The left and right
panels are the results corresponding to $\pi^++\pi^-$, $K^++K^-$,
and $p+\bar p$; as well as $\phi$ and $\it \Xi^-$, respectively.
One can see that the model results approximately consist with the
experimental data in special transverse momentum ranges in small
collision system measured at the LHC by the ALICE and CMS
Collaborations. The special transverse momentum ranges in $p$-Pb
collisions are not obviously observed in the available data
ranges. In $pp$ collisions, some particles are expected to have a
special transverse momentum range in 0--3.5 GeV/$c$ or a little
more.

\begin{figure*}[!htb]
\begin{center}
\includegraphics[width=16.cm]{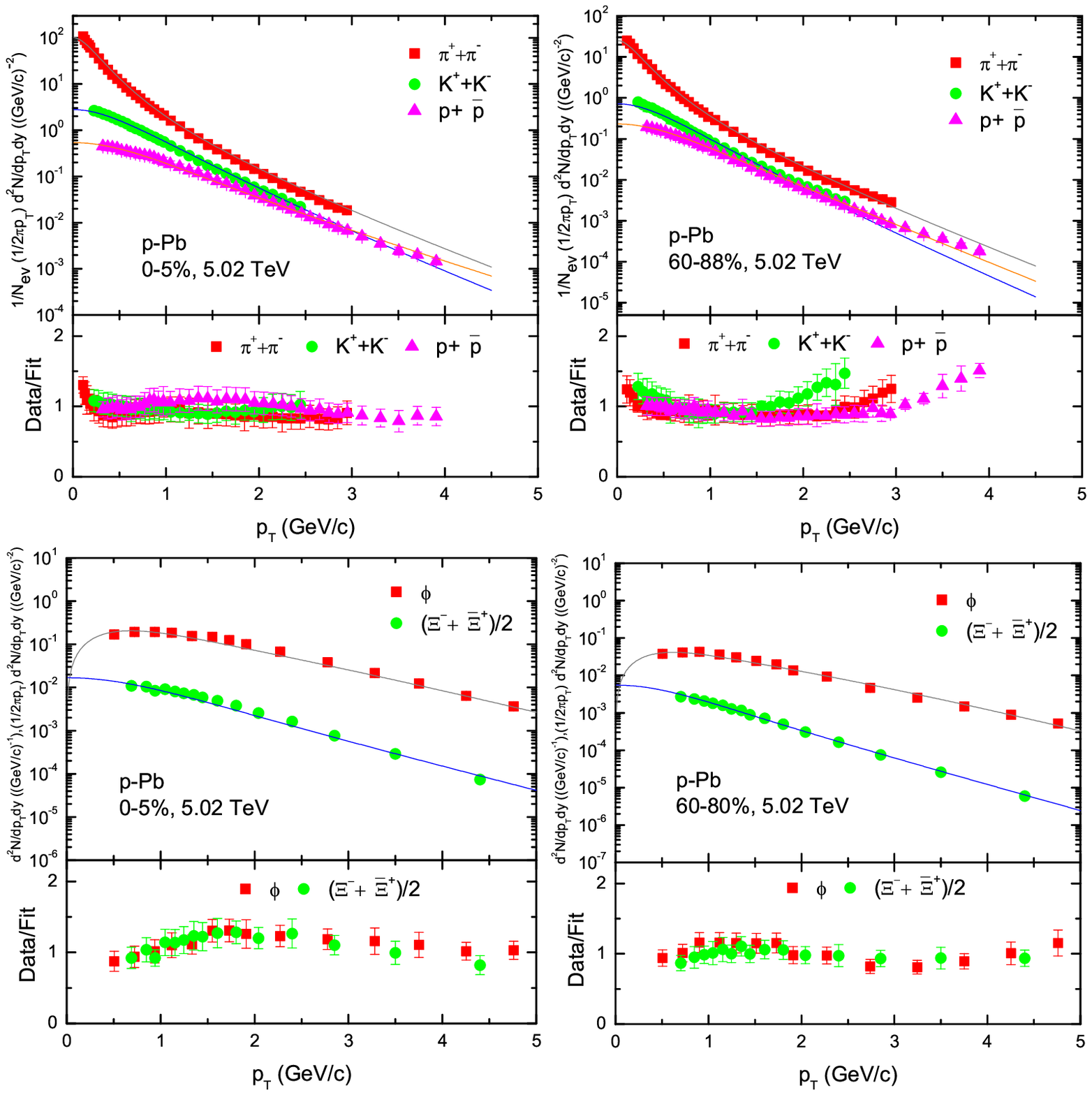}
\end{center}
Fig. 5. Same as Fig. 4, but showing the spectra in $p$-Pb
collisions at $\sqrt{s_{NN}}=5.02$ TeV, where different expression
of $\phi$ spectra is used. The symbols represent the experimental
data measured by the ALICE Collaboration [23--25], where the
spectra of $\phi$ and $\it(\Xi^-+\bar\Xi^+)/$2 are cut at 5
GeV/$c$. Left: for central collisions (0--5\% centrality); Right:
for peripheral collisions (70--80\% centrality). Upper: for
$\pi^++\pi^-$, $K^++K^-$, and $p+\bar p$ with $0<y<0.5$ [23];
Lower: for $\phi$ [24] and $\it(\Xi^-+\bar\Xi^+)/$2 [25] with
$-0.5<y<0$.
\end{figure*}

\begin{figure*}[!htb]
\begin{center}
\includegraphics[width=16.cm]{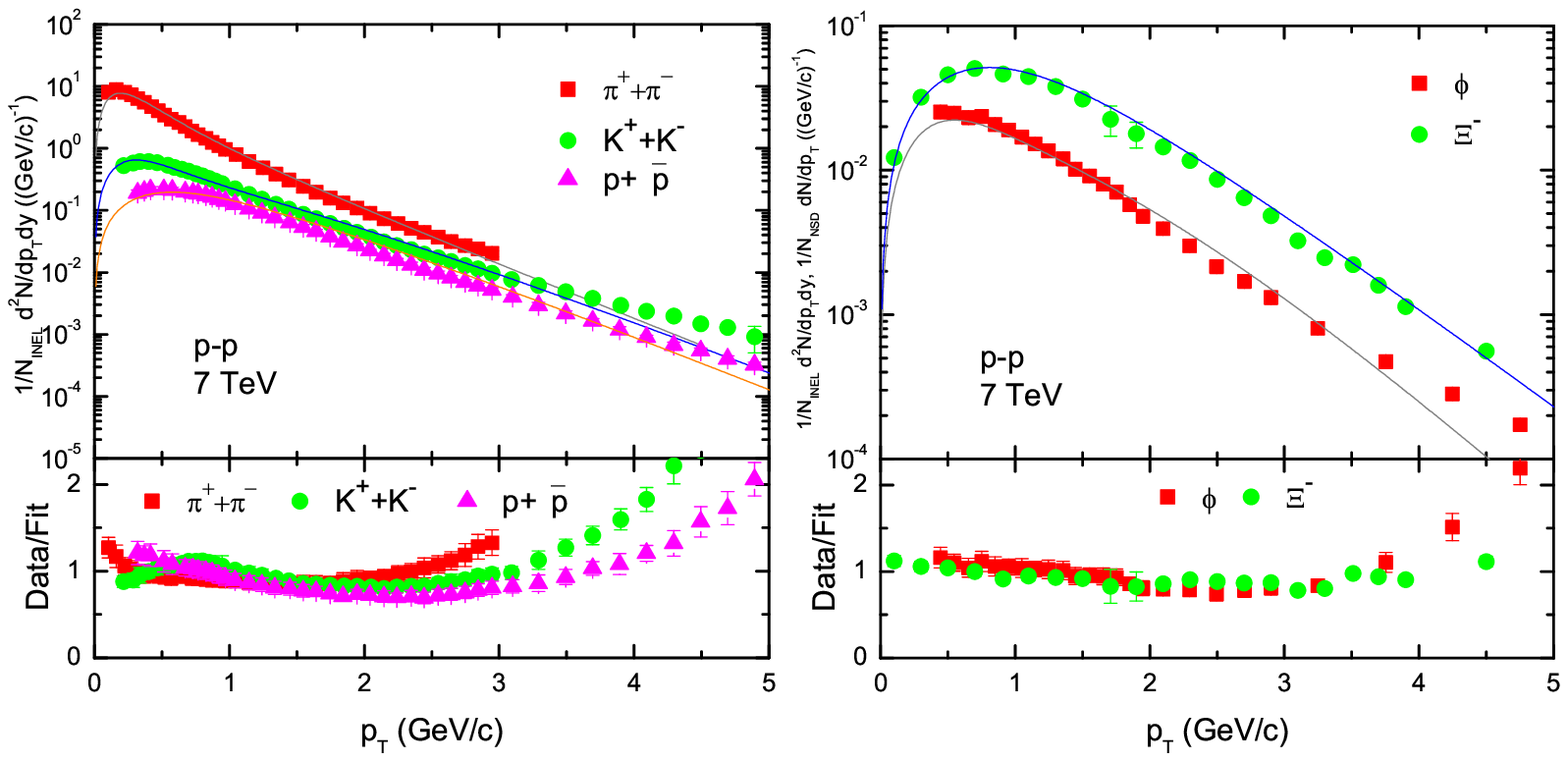}
\end{center}
Fig. 6. Same as Fig. 4, but showing the spectra with different
expressions in $pp$ collisions at $\sqrt{s}=7$ TeV, where $N_{\rm
INEL}$ and $N_{\rm NSD}$ denote the numbers of INEL and NSD events
respectively. The symbols represent the experimental data measured
by the ALICE [26, 27] and CMS [28] Collaborations, where the
spectra of $K^++K^-$, $p+\bar p$, $\phi$, and $\it \Xi^-$ are cut
at 5 GeV/$c$. Left: for $\pi^++\pi^-$, $K^++K^-$, and $p+\bar p$
with $|y|<0.5$ in INEL events [26]; Right: for $\phi$ with
$|y|<0.5$ in INEL events [27] and $\it \Xi^-$ with $|y|<2$ in NSD
events [28].
\end{figure*}

To study the change trends of free parameters with rest mass of
particle, Figure 7 gives the dependence of $T_0$ on $m_0$ (upper
panel) and the dependence of $\beta_T$ on $m_0$ (lower panel). The
left panels are for central and peripheral Au-Au collisions,
central and peripheral $d$-Au collisions, and $pp$ collisions at
200 GeV. The right panels are for central and peripheral Pb-Pb
collisions at 2.76 TeV, central and peripheral $p$-Pb collisions
at 5.02 TeV, and $pp$ collisions at 7 TeV. Different symbols
represent values of parameters in different collisions from Table
1. One can see that $T_0$ ($\beta_T$) increases (decreases)
slightly with the increase of $m_0$. $T_0$ ($\beta_T$) in central
collisions is slightly larger than or nearly equal to that in
peripheral collisions. $T_0$ ($\beta_T$) in collisions at the LHC
is slightly larger than or nearly equal to that in collisions at
the RHIC. $pp$ collisions are closer to peripheral nuclear
collisions.

\begin{figure*}[!htb]
\begin{center}
\includegraphics[width=16.cm]{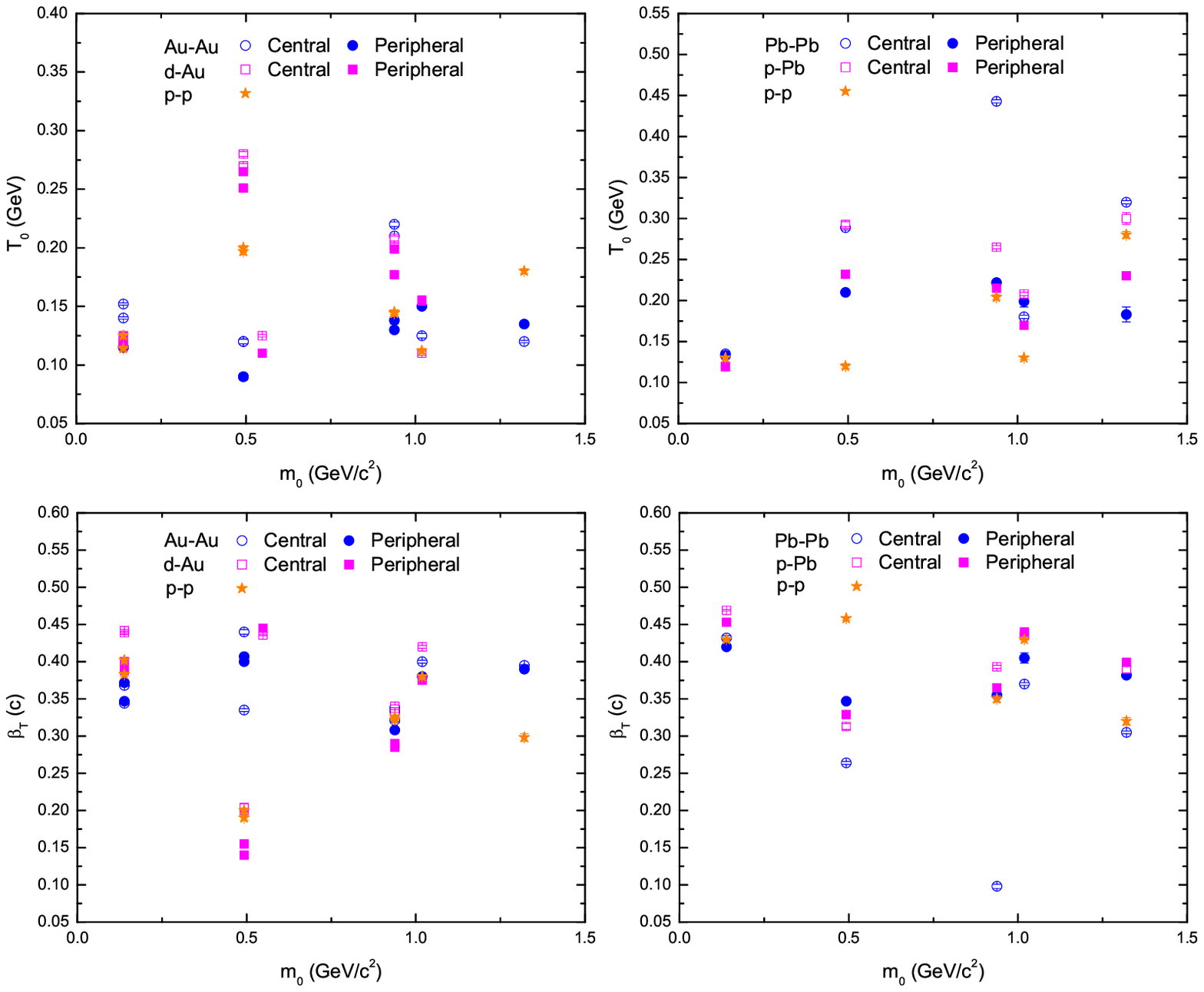}
\end{center}
Fig. 7. Upper: dependence of $T_0$ on $m_0$; Lower: dependence of
$\beta_T$ on $m_0$. Left: for central and peripheral Au-Au
collisions, central and peripheral $d$-Au collisions, and INEL or
NSD $pp$ collisions at 200 GeV; Right: for central and peripheral
Pb-Pb collisions at 2.76 TeV, central and peripheral $p$-Pb
collisions at 5.02 TeV, and INEL or NSD $pp$ collisions at 7 TeV.
Different symbols represent values of parameters in different
collisions, which are taken from Table 1.
\end{figure*}

The fact that $T_0$ is dependent of $m_0$ reveals the scenario for
multiple kinetic freeze-out at the RHIC and LHC [35, 56, 57]. The
scenario for single or double kinetic freeze-out [29, 58, 59] is
not observed in this study. Just as its name implies, the scenario
for multiple kinetic freeze-out uses different sets of parameters
for the spectra of different particles with different masses,
i.e., this scenario is the mass-dependent and differential. The
scenario for single kinetic freeze-out uses one set of parameters
for the spectra of all particles which includes both the strange
and non-strange particles. The scenario for double kinetic
freeze-out uses different sets of parameters for strange and
non-strange particles. Although the average values of parameters
weighted by different particles yield in multiple kinetic
freeze-out scenario can be regarded as the parameters in single
kinetic freeze-out scenario, the fit results will be unacceptable
due to large departure from the data.

In fact, if we use average $T_0$ and $\beta_T$ as the parameters
for single freeze-out scenario, the fitting results for pion
spectra are approximately acceptable due to these parameters being
close to those for pion emission. The fitting results for kaon and
proton spectra are not acceptable due to very large $\chi^2$ (in
some cases $\chi^2>1000$). In the case of double freeze-out
scenario, the average parameters weighted from the pion and proton
spectra can fit approximately the pion spectra, and cannot fit the
proton spectra, though the kaon spectra are described by another
set of parameters. Even if we fit the spectra in a narrower $p_T$
range, the fitting results by the single and double freeze-out
scenarios are not acceptable. In view of this situation, the
present work does not support the single and double freeze-out
scenarios, but the multiple kinetic freeze-out scenario.

The relative sizes of $T_0$ ($\beta_T$) in central and peripheral
nuclear collisions, as well as in collisions at the RHIC and LHC
are harmonious to our recent works [5, 60] in general, though the
contribution of soft excitation is included and the contribution
of hard scattering processes is excluded in low $p_T$ region in
[5]. Oppositely, the present work includes the contributions of
soft excitation process and hard scattering process in low $p_T$
region [60] in the Hagedorn's model [Eq. (4)], if the contribution
of hard process in low $p_T$ region is available. This causes that
there are some small differences in absolute values from our
previous work [5]. More discussions on the relative sizes of $T_0$
($\beta_T$) and comparisons with other works can be found in [60].

Although this work presents two superpositions, i.e. Eqs. (3) and
(4), they are not directly used to fit the data in Figs. 1--6.
Instead, Eq. (1) is simply used to fit the mentioned data. This
treatment uses in fact only the first component in Eqs. (3) and
(4), and it is more close to Eq. (4). In Eq. (4), the contribution
of hard process in low $p_T$ region is not excluded, though it
does not get entanglement in the determination of parameters of
the two components. While, Eq. (3) includes the contribution of
hard process in low $p_T$ region, and it gets entanglement in the
determination of parameters. To show main topic, the present work
gives up to use Eqs. (3) and (4) in high $p_T$ region beyond 5
GeV/$c$, though most data are only in low $p_T$ region. Instead,
Eqs. (3) and (4) are used in our recent works [5] and [60]
respectively, in which the functional trends in high $p_T$ region
can be clearly seen.

In our opinion, Eqs. (3) and (4) can be used to quantify the
centrality, rapidity, and energy dependent soft versus hard
contributions to particle production. This would bring some new
idea to business. For example, from the comparisons between the
curves and data in Figs. 1--3, we can see that the fraction of
hard contribution in peripheral collisions is greater than that in
central collisions. The fraction of hard contribution in $pp$
collisions is greater than that in central $d$-Au collisions, and
the latter is greater than that in central Au-Au collisions. The
reason is less cascade or intra-nuclear collisions in peripheral
collisions and in small system. Generally, the fractions in these
collisions are about a few percentages to above ten percentages
[5, 60]. To make a definitive conclusion on other dependences,
more and systematic data are needed. This is beyond the focus of
the present work. We shall not discuss them anymore.

The values of average $T_0$ ($\langle T_0\rangle$) and average
$\beta_T$ ($\langle \beta_T\rangle$) for different types of
collisions at the RHIC and LHC are listed in Table 2. These
average values are obtained by different weights due to different
yields ($N_0$) of $\pi^++\pi^-$, $K^++K^-$, and $p+\bar p$. Other
particles are not included in the averages due to non-identity
type and centrality. In particular, $\langle T_0\rangle$ in
central nuclear collisions is $\sim148$ MeV which is less than
$T_{ch}$ ($\sim160$ MeV [1--4]), which renders that the kinetic
freeze-out in central collisions at the considered energies
happens later than the chemical freeze-out by $\sim2$ fm according
to the time evolution of temperature, $T_f= T_i
(\tau_i/\tau_f)^{1/3}$ [61, 62], where $T_i$ ($=300$ MeV) and
$\tau_i$ ($=1$ fm) are the initial temperature and proper time
respectively [62]. Eq. (4) used in our recent work [60] results in
larger $T_0$ and $\beta_T$ than Eq. (3) used in our another work
[5] due to the non-exclusion of hard process in Eq. (4).

\begin{table*}
{\small Table 2. Values of $\langle T_0\rangle$ and $\langle
\beta_T\rangle$ in different types of collisions at the RHIC and
LHC. The average values are obtained by different weights due to
different yields ($N_0$) of $\pi^++\pi^-$, $K^++K^-$, and $p+\bar
p$. Other particles are not included in the averages due to
non-identity type and centrality. \vspace{-.30cm}
\begin{center}
\begin{tabular}{cccc}\\ \hline\hline
Collisions & Energy & $\langle T_0\rangle$ (GeV) & $\langle
\beta_T\rangle$ ($c$) \\ \hline
Central Au-Au     & 200 GeV  & $0.145\pm0.009$ & $0.366\pm0.018$ \\
Peripheral Au-Au  & 200 GeV  & $0.113\pm0.006$ & $0.363\pm0.013$ \\
Central $d$-Au    & 200 GeV  & $0.140\pm0.013$ & $0.411\pm0.045$ \\
Peripheral $d$-Au & 200 GeV  & $0.135\pm0.013$ & $0.369\pm0.041$ \\
INEL or NSD $pp$  & 200 GeV  & $0.128\pm0.006$ & $0.371\pm0.017$ \\
\hline
Central Pb-Pb     & 2.76 TeV & $0.164\pm0.014$ & $0.410\pm0.037$ \\
Peripheral Pb-Pb  & 2.76 TeV & $0.146\pm0.009$ & $0.409\pm0.025$ \\
Central $p$-Pb    & 5.02 TeV & $0.145\pm0.002$ & $0.448\pm0.004$ \\
Peripheral $p$-Pb & 5.02 TeV & $0.137\pm0.011$ & $0.435\pm0.033$ \\
INEL or NSD $pp$  & 7 TeV    & $0.133\pm0.005$ & $0.429\pm0.012$ \\
\hline
\end{tabular}%
\end{center}}
\end{table*}

In different types of collisions, the values of $T_0$ and
$\beta_T$ for emissions of pions show small fluctuations due to
large yields and high statistics. The values of $T_0$ and
$\beta_T$ for emissions of other particles show large fluctuations
due to small yields and low statistics. Because of large yields of
pions, $\langle T_0\rangle$ and $\langle \beta_0\rangle$ weighted
for yields of different particles are closer to those for
emissions of poins. Since the emissions of other particles require
higher $T_0$, other particles freeze-out earlier than pions at the
kinetic freeze-out stage. To study accurately dependences of $T_0$
($\beta_T$) on centrality and energy, one can select the spectra
of pions, at the most including the spectra of kaons and
(anti-)protons. In fact, in the alternative method discussed in
the beginning of section 2, the spectra of the three types of
particles are needed.

Note that the extracted mass dependent $T_0$ and $\beta_T$ in
central collisions are comparable with those in peripheral
collisions. In particular, the behavior of $\beta_T$ is different
from that of the elliptic flow $v_2$ because $\beta_T$ describes
the flow velocity in the transverse plane and $v_2$ describes the
flow anisotropy in the transverse plane. In central collisions,
$v_2$ is approximately equal to 0 due to the approximately
isotropy. While in peripheral collisions, $v_2$ is considerable
due to the anisotropy. In some cases, the velocity may be small,
and the anisotropy may be large. The mass dependent parameters may
be related to the scaling of the number of constituent quarks. The
concrete relation is worth to be studied in the future.

We would like to point out that the errors of parameters are very
small due to strict restrictions ($\chi^2_{\max} \leq
1.05\chi^2_{\min}$) being used in the fits (where $\chi^2_{\max}$
denotes the maximum-$\chi^2$ when we determine the errors and
$\chi^2_{\min}$ denotes the minimum-$\chi^2$ when we determine the
best parameters). If we use weak restrictions, large errors will
be obtained. However, in this case, large $\chi^2$ will also be
obtained, which shows bad fitting results. On the other hand, as a
fitting function itself, Eq. (4) is not ideal due to fewer
parameters being used in low $p_T$ region, which renders small
variable ranges of parameters in limited selection. Instead, Eq.
(3) can give better fit due to more parameters being used in low
$p_T$ region, which renders large variable ranges of parameters by
flexible selection. Indeed, the limited selection restricts $T_0$
and $\beta_T$ themselves.

Before the final conclusions, it should be emphasized that the
successful fitting of the blast-wave model for the experimental
data studied reveals that the properties of interacting system and
produced particles at the kinetic stage is abundant. The
excitation degree of interacting system and the collective motion
of produced particles are described in the model by $T_0$ and
$\beta_T$ respectively. According to $T_0$ and $\beta_T$ for the
emissions of different particles, we can study various possible
scenarios which includes single, double, or multiple kinetic
freeze-out. The present work conforms the multiple kinetic
freeze-out scenario. Although the blast-wave model is not the only
option in the extraction of $T_0$ and $\beta_T$, it is one of the
applicable methods.

In particular, the blast-wave model makes an assumption that
particles are locally thermalized at a kinetic freeze-out
temperature and are moving with a common collective transverse
radial flow velocity field. It seems that a single freeze-out
scenario, that is the same $T_0$ and $\beta_T$, should be taken
into consideration for different particles. The present work shows
that $T_0$ and $\beta_T$ for the emissions of different particles
are severally different. This means that the interacting system is
locally thermalzied or successionally emissive in various types of
collisions. Even in peripheral nucleus-nucleus collisions and in
$pp$ collisions, the multiplicity per unit rapidity at the
mid-rapidity is enough high at the RHIC and LHC, and the concept
of equilibrium statistical mechanics can be used.
\\

{\section{Conclusions}}

We summarize here our main observations and conclusions.

(a) The transverse momentum spectra of different types of
particles produced in mid-rapidity interval in central and
peripheral Au-Au collisions at 200 GeV, central and peripheral
$d$-Au collisions at 200 GeV, and INEL or NSD $pp$ collisions at
200 GeV, as well as in central and peripheral Pb-Pb collisions at
2.76 TeV, central and peripheral $p$-Pb collisions at 5.02 TeV,
and INEL or NSD $pp$ collisions at 7 TeV are analyzed by the
blast-wave model with Boltzmann-Gibbs statistics. In the model
fit, the contributions of soft excitation and hard scattering
processes are included in low transverse momentum region. This
treatment of Hagedorn's model leads to fewer parameters and
smaller errors of parameters in low transverse momentum region.
The model results are approximately in agreement with the
experimental data measured at the RHIC by the PHENIX and STAR
Collaborations and at the LHC by the ALICE and CMS Collaborations.

(b) The kinetic freeze-out temperature increases slightly and the
transverse flow velocity decreases slightly with the increase of
particle mass. The kinetic freeze-out temperature in central
collisions is slightly larger than or nearly equal to that in
peripheral collisions, and that in collisions at the LHC is
slightly larger than or nearly equal to that in collisions at the
RHIC. The dependences of transverse flow velocity on centrality
and energy are similar to those of the kinetic freeze-out
temperature. The similarity of $pp$ collisions to peripheral
collisions is observed by the Hagedorn's model in which the
contribution of hard process in low transverse momentum region is
included. The fact that the kinetic freeze-out temperature is
dependent on particle mass reveals the scenario for multiple
kinetic freeze-out at the RHIC and LHC. The scenario for single or
double kinetic freeze-out is not observed in this study.
\\
\\
{\bf Data Availability}

The data used to support the findings of this study are quoted
from the mentioned references. As a phenomenological work, this
paper does not report new data.
\\
\\
{\bf Conflicts of Interest}

The authors declare that there are no conflicts of interest
regarding the publication of this paper.
\\
\\
{\bf Acknowledgments}

Communications from Edward K. Sarkisyan-Grinbaum are highly
acknowledged. This work was supported by the National Natural
Science Foundation of China under Grant Nos. 11575103 and
11747319, the Chinese Government Scholarship (China Scholarship
Council), the Shanxi Provincial Natural Science Foundation under
Grant No. 201701D121005 (China), the Fund for Shanxi ``1331
Project" Key Subjects Construction (China), and the Grant of
Scientific Research Deanship at Qassim University (Kingdom of
Saudi Arabia).
\\
\\
\end{multicols}
\end{document}